\documentclass[
    paper,
  ]{geophysics}

\usepackage[textsize=footnotesize, textwidth=2.8cm]{todonotes}
\usepackage[utf8x]{inputenc}
\usepackage{lmodern}
\usepackage[T1]{fontenc}

\usepackage{amssymb, amsmath, amsfonts}
\usepackage{upquote}
\usepackage[strings]{underscore}
\usepackage{siunitx}                % SI conform commands (eg \num)
\usepackage{tabularx}
\usepackage{colortbl}
\usepackage{booktabs}
\usepackage{xspace}
\usepackage{flafter}

\usepackage[pdftex, final]{hyperref}
\hypersetup{allcolors=blue, allbordercolors={0 0 .5}, colorlinks=true}

\DeclareGraphicsExtensions{.pdf,.png,.jpg}

\ifthenelse{\boolean{@twoc}}{%
  }{%
  }

\ifthenelse{\boolean{@twoc}}{%
  }{%
  }

\begin{document}

\title{
Geophysical inversions to delineate rocks with CO$_2$ sequestration potential through carbon mineralization
}

\renewcommand{\thefootnote}{\fnsymbol{footnote}}

\ms{}  % ARTICLE-ID

\author{%
Lindsey J. Heagy\footnotemark[1],
Thibaut Astic\footnotemark[1],
Joseph Capriotti\footnotemark[1],
John Weis\footnotemark[1],
and Douglas W. Oldenburg\footnotemark[1], \\
\footnotemark[1]Department of Earth, Ocean and Atmospheric Sciences, University of British Columbia, Vancouver, British Columbia \\
e-mail: \href{mailto:lheagy@eoas.ubc.ca}{lheagy@eoas.ubc.ca}
}

\footer{}
\lefthead{}
\righthead{}

\maketitle

\begin{abstract}
  In addition to reducing anthropogenic emissions of CO$_2$, it is increasingly clear we also need to remove CO$_2$ from the atmosphere in order to avoid some of the worst case scenarios for climate change. Geologic sequestration of CO$_2$ is among the most attractive approaches because of the large global capacity and long-time scales for storage. One mechanism of geologic storage is through carbon mineralization. Some mafic and ultramafic rocks contain minerals that will react with CO$_2$ in a carbonation reaction and convert it to carbonated minerals. This is effectively a permanent CO$_2$ storage mechanism. The geologic question we are faced with is then to locate, delineate and estimate the volume of potentially reactive rocks. Using a synthetic model that emulates a prospective site for carbon mineralization in British Columbia, we simulate and invert gravity and magnetic data to delineate reactive rocks. We begin by inverting each data set independently and introduce a proxy experiment to contend with the challenging problem of choosing an appropriate physical-property threshold to estimate volumes from the recovered model. We use this proxy experiment to estimate thresholds for standard, $\ell_2$ inversion of the gravity and magnetics, as well as for inversions which use sparse and compact norms. A Petrophysically and Geologically Guided Inversion (PGI) framework is used to construct quasi-geologic models from which volumes can be estimated directly. We apply the PGI framework to the magnetics and gravity data independently. The framework is also used to jointly invert these data and produce a model that is consistent with both data sets. Cumulative volume estimates with depth are informative and can help decide whether in situ or ex situ sequestration might be appropriate. Using each of the inverted models, we estimate cumulative volume of reactive rock as a function of depth.
\end{abstract}

\section{Introduction}
Even if we are successful in reaching net-zero carbon emissions by 2050, the recent IPCC report predicts that global warming will exceed 1.5°C in the 21st century \citep{IPCC2021}. This conclusion highlights that while it is important to transition off of hydrocarbon-based energy sources, it is also critical that we remove CO$_2$ from the atmosphere. Since 1850, over 2000 GtCO$_2$ of anthropogenic CO$_2$ have been emitted \citep{IEA2021}, and despite the COVID-19 pandemic, energy-related emissions in 2020 energy-related emissions were 31.5 GtCO$_2$. Thus, we are faced with the need for negative emissions technologies (NETs),  also referred to as carbon dioxide removal (CDR) methods, to remove hundreds to thousands of Gt of CO$_2$ from the atmosphere \citep{IPCC2021, NationalAcademies2019}. The 2019 report from the National Academies \citep{NationalAcademies2019} provides an overview of 6 categories of approaches to capture and store CO$_2$. Of all of the approaches presented, geologic storage is the most desirable because of both the long time scales of storage as well as the global capacity. Two mechanisms involve geologic storage: sequestration of CO$_2$ in sedimentary formations and carbon mineralization of CO$_2$.

The first mechanism is the more mature of the two and involves the injection of supercritical CO$_2$ into depleted hydrocarbon reservoirs or deep saline aquifers \citep{NationalAcademies2019, Kelemen2019}. This approach relies on there being a porous, permeable reservoir overlain by a caprock that prevents CO$_2$ migration out of the reservoir. Major risks for CO$_2$ migration include non-sealed fractures in the caprock and leaky wellbores. Geophysical data, including seismic, InSAR, and electromagnetics can be employed to monitor wellbore integrity, injection, and leakage; there are multiple field examples worldwide (e.g. \cite{Vasco2010, Ajo-Franklin2013, Heagy2019, Wilt2020}).

The second mechanism for geologic storage is carbon mineralization; this is the focus of this paper. This approach utilizes magnesium and calcium-bearing minerals in mafic or ultramafic rocks that react with CO$_2$ and convert it to solid, stable, carbonate minerals \citep{NationalAcademies2019, Seifritz1990, Lackner1995}. Although the science and technology of carbon mineralization are less mature than geologic storage in sedimentary units, this approach presents several potential advantages. Storage is permanent with minimal risk of leakage \citep{Zhang2017}, and it is estimated that there may be orders of magnitude more global capacity than sequestration in sedimentary units \citep{Kelemen2019}. Broadly speaking, there are two strategies for promoting reaction with CO$_2$: (a) in-situ approaches which require circulating CO$_2$-bearing fluids through subsurface formations where they can react, and (b) ex-situ (or surficial) approaches use minerals that have been brought to the surface, for example in mine-tailings \citep{NationalAcademies2019, Kelemen2019}. In some places, reactive rocks are co-located with minerals of economic interest. At these locations, it may be possible for mines to have net-zero, or even net negative, emissions and ``carbon mines'' that would bring reactive minerals to the surface might become economically viable. For in-situ approaches, there are only two pilot projects that have been conducted: the Carbfix Project in Iceland \citep{Gislason2018} and the Wallula Project in Washington State \citep{McGrail2014}. Both of these sites are injections into basalts. To the best of our knowledge, there have not yet been in-situ experiments in ultramafic rocks.

For either an in-situ or ex-situ approach, the first step is to identify the location and volume of reactive rocks. Recent work in \citep{Cutts2021, Mitchinson2020} examines the physical properties of ultramafic rocks as they are altered. Of particular interest are rocks that have been altered through serpentinization and can react with CO$_2$ through a carbonation reaction. As ultramafic rocks go through each reaction phase, their magnetic susceptibility and density are altered. This motivates the use of geophysical data to locate these rocks, estimate their depth and volume, and potentially even estimate their reactivity.

In this paper, we will identify some of the opportunities and areas of future research for the use of geophysical data for carbon mineralization. We will focus on ultramafic rocks using the physical property work in \citep{Cutts2021} and the initial inversion of magnetic data over ultramafic sites in BC \citep{Mitchinson2020} as motivation. The paper is organized as follows. First, we will discuss how the physical properties of ultramafic rocks change as they are altered. We then introduce a representative model and simulate gravity and magnetic data. Next, we will introduce approaches for independently and jointly inverting these data. Finally, we will conclude with a discussion of some of the practical challenges and research opportunities for the use of geophysical data for carbon mineralization. All of the code used to generate figures for this paper is available at https://github.com/simpeg-research/Fast-Times-2021-Carbon-Mineralization.

\section{Physical Properties}
In the context of ultramafic rocks, rocks that have been serpentinized are of particular interest for carbon mineralization because they are particularly reactive with CO$_2$. Following the discussion in \cite{Cutts2021}, the general formula for a serpentinization reaction can be summarized by:

\begin{displaymath}
    \text{R1:}~ \text{olivine} \pm \text{orthopyroxene} + \text{H}_2\text{O}
    \rightarrow
    \text{serpentine} \pm \text{brucite} \pm \text{magnetite}
    \label{eq:serpentinization}
\end{displaymath}

Rocks that have been serpentinized are then susceptible to reaction with CO$_2$ through carbonation reactions. Brucite is of particular interest for carbon mineralization because it reacts rapidly with CO$_2$ at surface conditions \citep{Kelemen2019}. The general formulae for carbonation reactions are \citep{Hansen2005}

\begin{displaymath}
    \begin{split}
    \text{R2:}~ \text{olivine} + \text{brucite} + \text{CO}_2 + \text{H}_2\text{O}
    &\rightarrow
    \text{serpentine} + \text{magnesite} + \text{H}_2\text{O}
    \\
    \text{R3:}~ \text{serpentine} + \text{CO}_2
    &\rightarrow
    \text{magnesite} + \text{talc} + \text{H}_2\text{O}
    \\
    \text{R4:}~ \text{talc} + \text{CO}_2
    &\rightarrow
    \text{magnesite} + \text{quartz} + \text{H}_2\text{O}
    \end{split}
    \label{eq:carbonation}
\end{displaymath}

The key connection point between any question about the subsurface and the use of geophysical data is physical properties. Serpentinization and carbonation reactions change both the density and magnetic susceptibility of rocks. The density and magnetic susceptibility of fresh ultramafic rocks should be similar to that of olivine and pyroxene (3.10-3.30 g/cc and $~10^{-3}$ SI). Minerals produced through serpentinization (serpentine, brucite and magnetite) are typically less dense than fresh ultramafics. Brucite and serpentine have densities of approximately 2.57 and 2.39 g/cc, respectively. Magnetite also has a larger magnetic susceptibility than unaltered phases. So overall, serpentinized rocks have a lower density and higher magnetic susceptibility than fresh ultramafics. Brucite, serpentine and magnetite are consumed in a carbonation reaction forming magnesite as well as talc and quartz. Magnesite has a density of approximately 3.00 g/cc, and thus carbonated rocks are more dense and have a lower magnetic susceptibility than serpentinites \citep{Cutts2021, Hansen2005}.

\begin{figure}[!htb]
    \begin{center}
    \includegraphics[width=0.9\textwidth]{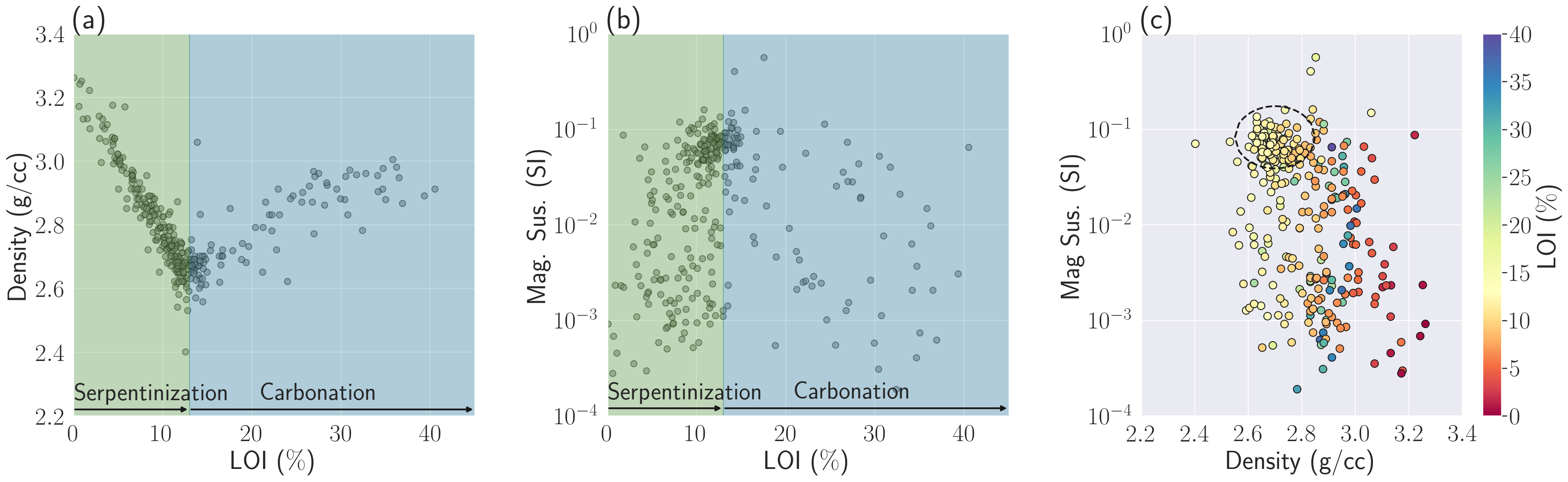}
    \end{center}
\caption{
    Physical property measurements of (a) density and (b) magnetic susceptibility of ultramafic rocks as a function of Loss On Ignition from \cite{Cutts2021}. From an LOI of 0\% to $\sim$13\%, rocks are undergoing serpentinization. Beyond an LOI of $~\sim$13\%, rock undergo carbonation. (c) Cross plot of density and magnetic susceptibility colored by Loss On Ignition (LOI).
}
\label{fig:phys-props-loi}
\end{figure}

To quantify how physical properties of ultramafic rocks change as a function of alteration, \cite{Cutts2021} made geochemical and petrophysical measurements on over 400 samples of ultramafic rocks from the Canadian Cordillera. In Figure \ref{fig:phys-props-loi}, we reproduce portions of Figure 6 in their paper and show physical property measurements as a function of Loss On Ignition (LOI), which is a proxy for alteration. From an LOI of 0\% to $\sim$13\%, fresh ultramafic rocks undergo a serpentinization reaction where fresh ultramafics are hydrated with the addition of water. Beyond an LOI of $\sim$13\%, magnesium and calcium-rich minerals react with CO$_2$ in carbonation reactions. Rocks that have been serpentinized but not yet carbonated have the most CO$_2$ uptake capacity and therefore are of the most interest for carbon mineralization. Figure \ref{fig:phys-props-loi} shows that (a) density and (b) magnetic susceptibility vary as a function of LOI. The trend with density is very clear: as fresh ultramafic rocks are serpentinized, the density decreases from $\sim$3.2 g/cc to $\sim$2.7 g/cc. Magnetic susceptibility is much more variable, but broadly speaking, serpentinized rocks are more susceptible than fresh ultramafic or carbonated rocks. Remanent magnetization is also important to consider in these settings. Though also quite variable, serpentinized rocks can have a substantial remanent magnetization, with many samples in the data collected by \cite{Cutts2021} having Koenigsberger ratios larger than 1 (see figures 6 and 8 in \cite{Cutts2021}). In the examples we develop in this paper, we will not include remanence. In practice, it will be important to consider \citep{Mitchinson2020}, and inversion approaches that enable us to recover the magnetization vector such as developed in \cite{Fournier2020} will be valuable tools.

Figure  \ref{fig:phys-props-loi} (c) shows the same physical property information as a cross-plot of density and magnetic susceptibility with the points colored by LOI. The orange yellow points are rocks that have been serpentinized but not yet significantly carbonated. Cross-plots are generally insightful and often yield well-defined clusters in physical property space that are associated with particular rock types. Although this is not easily visible here, there is a concentration of samples that have a high magnetic susceptibility, low density, and within the  desired LOI range for reactive serpentinized rocks.

\section{Motivating Example}
For both in-situ or ex-situ approaches to carbon mineralization, the goal of using geophysical data is to locate and delineate potentially reactive rocks. Important factors to characterize include: (a) the depth, which determines whether an in-situ or ex-situ approach should be used, (b) the volume, which influences the CO$_2$ sequestration capacity, and (c) for ultramafic rocks, the degree of serpentinization or carbonation which controls how reactive the rocks are.

\begin{figure}[!htb]
    \begin{center}
    \includegraphics[width=0.6\textwidth]{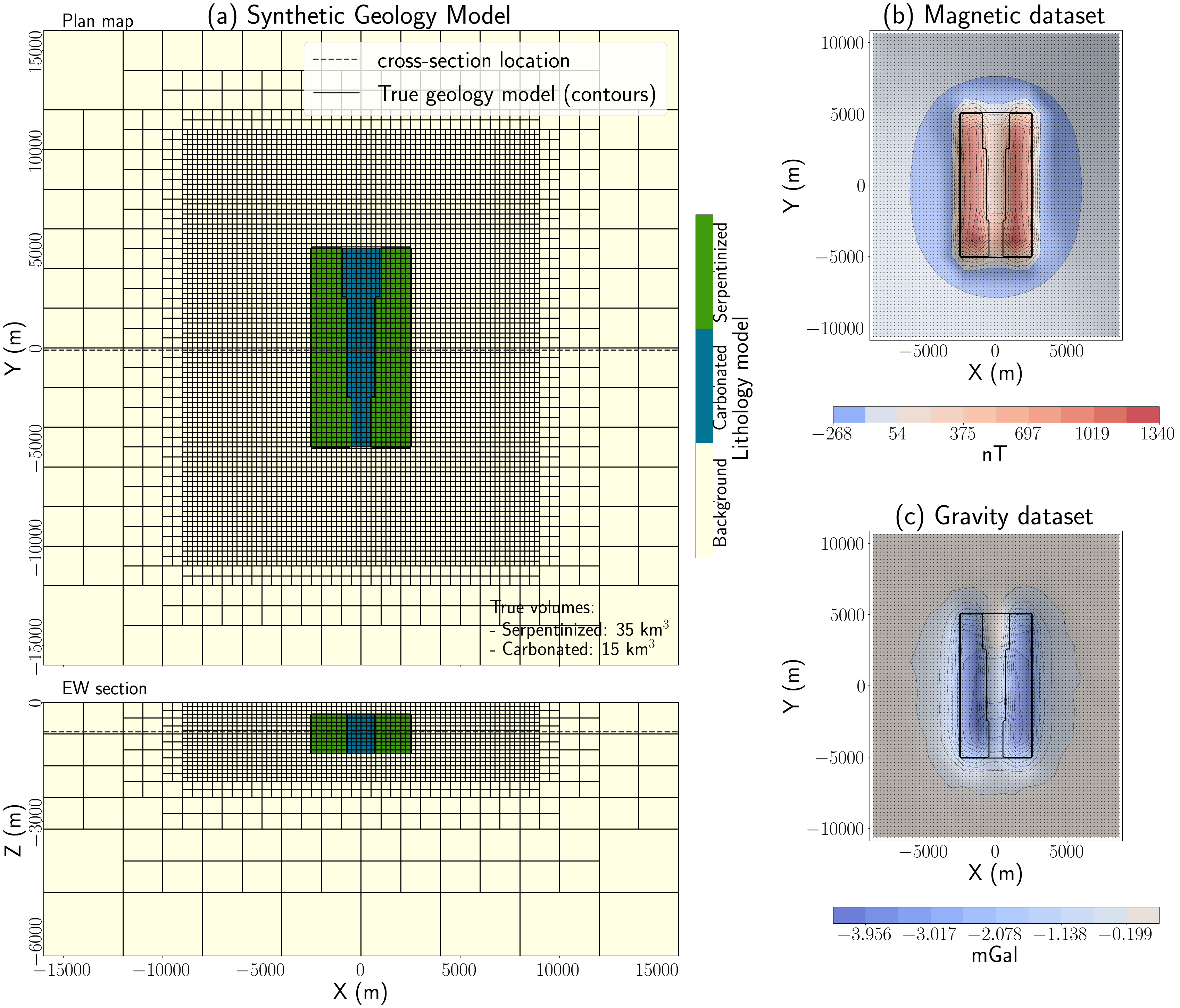}
    \end{center}
\caption{
    (a) Representative model of a serpentinized rock unit that contains a center region that has been carbonated. The physical properties assigned to each unit are shown in table \ref{tab:physical-properties}. The total volume of serpentinized rock is 35 km$^3$ and carbonated rock is 15 km$^3$. Panels show the simulated (b) magnetic and (c) gravity data.
}
\label{fig:representative-model}
\end{figure}

To illustrate the role of geophysical data and avenues for future development, we introduce a simple representative model in Figure \ref{fig:representative-model}a. This model is motivated by the geologic setting at the Decar site (e.g. see the maps and data in \cite{Mitchinson2020}). We summarize the physical properties of each unit in Table \ref{tab:physical-properties}. A serpentinized block with a magnetic susceptibility of 0.15 SI and a density of 2.7 g/cc is embedded in a halfspace background (0 SI, 2.9 g/cc: the background is composed of mafic volcanic and sedimentary rocks). Within that block, there is a region that has been carbonate-altered, it is wider in the north (2km) and tapers to the south (1km). The carbonated region has a magnetic susceptibility of 0.05 SI and a density of 3.0 g/cc.

\begin{table}
    \centering
        \caption{Physical properties of the rock units shown in Figure \ref{fig:representative-model}}
        \begin{tabular}[htb]{| l | p{3.7cm} | p{3.7cm}  | p{3.7cm}  |}
            \hline
             & \textbf{Magnetic susceptibility (SI)} & \textbf{Density (g/cc)} & \textbf{Density contrast (g/cc)}\\
            \hline
            background & $0$ & 2.9 & 0 \\
            serpentinized rock & 0.15 & 2.7 & -0.2 \\
            carbonated rock & 0.05 & 3.0 & 0.1 \\
            \hline
        \end{tabular}
        \label{tab:physical-properties}
     \end{table}

We simulate the gravity anomaly and magnetic response (assuming a vertical inducing field) and show those simulated data in Figure \ref{fig:representative-model} b and c, respectively. The data spacing in x and y is 250m. Gaussian random noise of 0.05 mGal and 1 nT are added to the gravity and magnetic data, respectively. These simulations, and the inversions we show subsequently, were run using the open-source SimPEG software \citep{Cockett2015}.

\section{Individual inversions}
As a first step, we invert each data set independently. We use a deterministic approach to solve the inverse problem. We formulate the inversion as an optimization problem where we minimize an objective function that is comprised of a data misfit term and a model norm term \citep{Tikhonov1977}.

\begin{equation}
\underset{m}{\text{minimize}} ~ \phi(m) = \phi_d(m) + \beta\phi_m(m)
\label{eq:inverse-problem}
\end{equation}

The data misfit term is a measure of how well the data that are predicted given a model $\mathbf{m}$ reproduce the observed data. The model norm (or regularization) term is included to overcome the non-uniqueness of the inverse problem and can be used to incorporate geologic knowledge or assumptions (see \cite{Oldenburg2005} for a tutorial). A standard approach is to use a term of the form:

\begin{equation}
    \phi_m(m) =
    \alpha_s\int |m - m_{{\rm ref}}|^p dV
    +
    \alpha_x\int \left|\frac{dm}{dx}\right|^{q} dV
    +
    \alpha_y\int \left|\frac{dm}{dy}\right|^{q} dV
    +
    \alpha_z\int \left|\frac{dm}{dz}\right|^{q} dV
    \label{eq:model-norm}
\end{equation}

The first term is often referred to as the ``smallness'' term while the subsequent terms are referred to as first-order ``smoothness'' in the $x, y, z$ directions, respectively. Generally $p=q=2$ is implemented but $(p,q)<2$ can be used to promote sparse or compact structures in the inversion \citep{Fournier2019}. Throughout, we will use $\ell_2$ to indicate an inversion where $p=q=2$, and $\ell_{pq}$ to indicate an inversion where $p<2$ and / or $q<2$.

\subsection{Magnetics}
\begin{figure}[!htb]
    \begin{center}
    \includegraphics[width=0.7\textwidth]{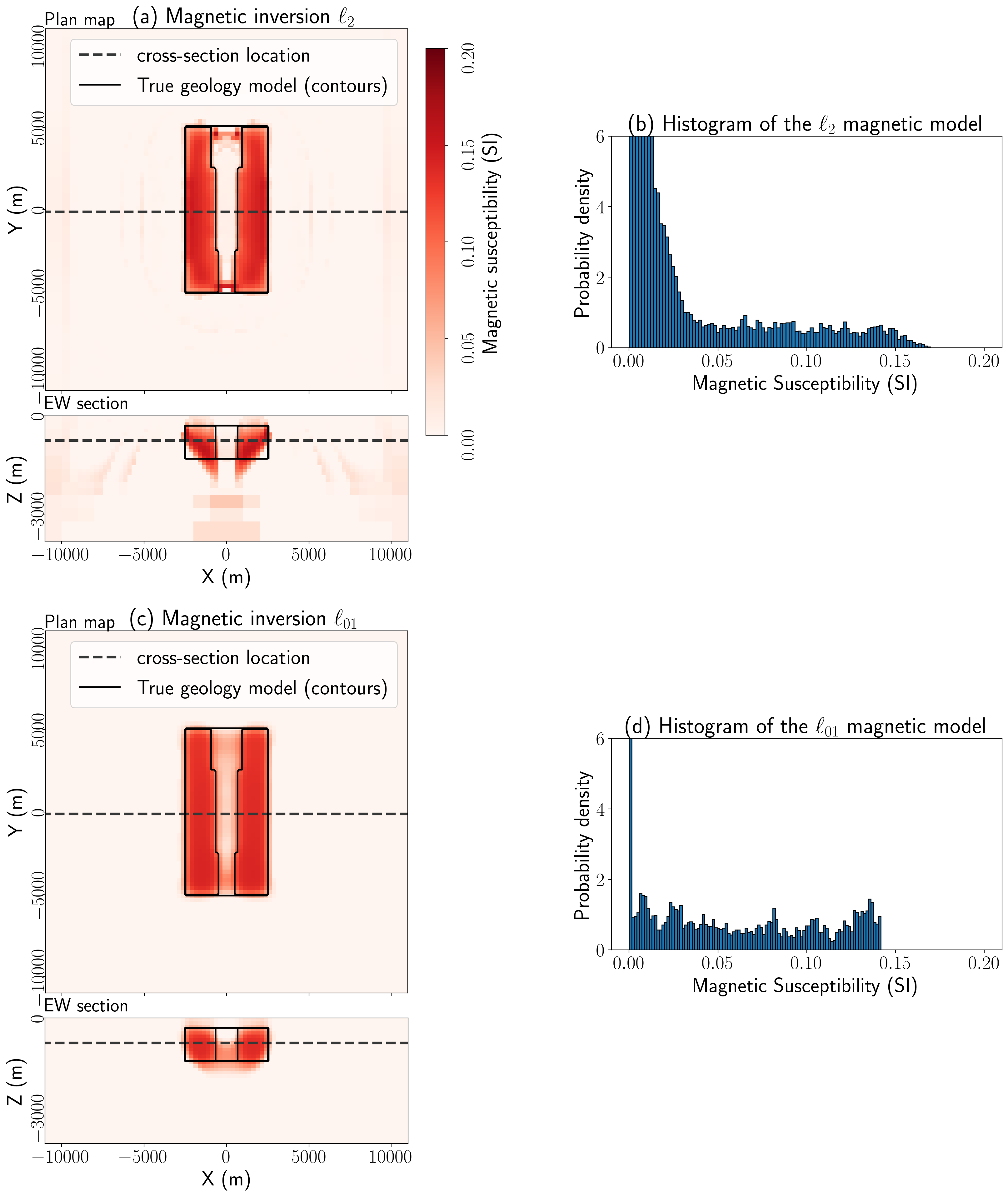}
    \end{center}
\caption{
    Models recovered using (a) an $\ell_2$ inversion of the magnetic data and (c) an $\ell_{01}$ inversion of the magnetic data. The top plot in each is a depth slice at $z=$-950m and the bottom panel shows a cross section along the line $y=0$m. Plots (b) and (d) show a histogram of the recovered susceptibility values. The histograms are normalized so that the area integrates to 1. Note that the y-limits clip the true maximum value in each.
}
\label{fig:magnetics-inversion}
\end{figure}

In Figure \ref{fig:magnetics-inversion}, we show inversion results for the magnetics data using (a) a standard $\ell_2$ inversion, and (c) a $\ell_{01}$ inversion. To compensate for the inherent lack of depth resolution of potential field data, we apply depth weighting in both inversions \cite{Li1996}. An $\ell_2$ inversion is the standard approach for inverting most data types and, in general, will favour models with smooth structures and variations in physical properties. There are some artefacts in the recovered model: there is some moderately susceptible material at depth and in the padding cells. We can also see the impacts of this regularization in the histogram of recovered susceptibility values in Figure \ref{fig:magnetics-inversion}(c) where there is a smooth distribution of physical property values with the maximum at 0 SI (the background susceptibility. We also see that very few cells reach the true susceptibility of the serpentinized unit (0.15 SI).

To promote sharper contrasts and more compact structures, we can alter the norms. In an $\ell_{pq}$ inversion, $\ell_2$ norms are used during the first stage of the optimization to produce a smooth initial model. Once the data are fit, the norms are then updated to allow for more compact structures and discontinuities in the final recovered model. Using $p=0$ and $q=1$, we recover the results shown in Figure \ref{fig:magnetics-inversion}c. In the histogram in Figure \ref{fig:magnetics-inversion}d, we see that there are many cells with the true background value, but this distribution is not ``smeared'’ to other low susceptibility values. Additionally, we see that the $\ell_{01}$ inversion recovers some moderate susceptibility values in the center, carbonated region, which in the true model has a susceptibility value of 0.05 SI.

\begin{figure}
    \begin{center}
    \includegraphics[width=0.5\textwidth]{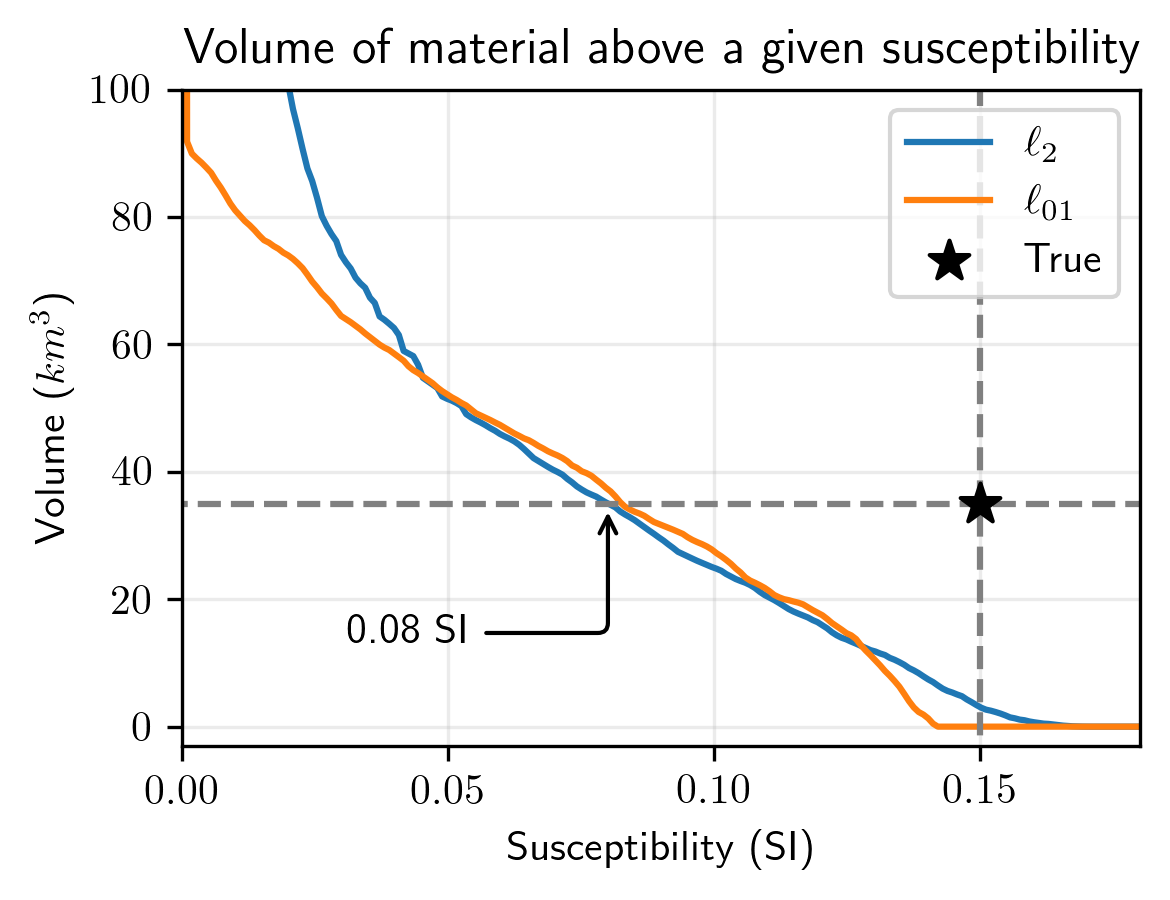}
    \end{center}
\caption{
    Estimated volume of serpentinized rock for a range of susceptibility thresholds.
}
\label{fig:magnetics-l2-lpq-volume}
\end{figure}

An important control on how much CO$_2$ can be sequestered is the total volume of serpentinized rock, and thus it is of interest to use these results to estimate the volume of serpentinized rock. In our synthetic model, the serpentinized rocks are distinguished from the carbonated rocks and background by having a high susceptibility (0.15 SI). Ideally, we would like to invert data and delineate the region of the model having this value. However, inversions generally produce models which are smoothed in some sense and hence very few cells in either the $\ell_2$ or $\ell_{01}$ inversions reach the true value of 0.15 SI. In the $\ell_{01}$ inversion, the maximum recovered susceptibility is 0.14 SI, so using the true value of 0.15 SI would result in a null volume. There is no clear way to choose a cut-off when looking at the physical property histograms in Figure \ref{fig:magnetics-inversion}. To examine how our choice of threshold impacts our estimate of volume of serpentinized rock, we plot estimated volume as a function of threshold for both inversion results in Figure \ref{fig:magnetics-l2-lpq-volume}. For each chosen threshold value, we compute the volume of material with a susceptibility larger than or equal to that value. We confine the region where we are developing an estimate to the x-y extent of the survey, and within a maximum depth of 3.5km.

\begin{figure}[!htb]
    \begin{center}
    \includegraphics[width=1\textwidth]{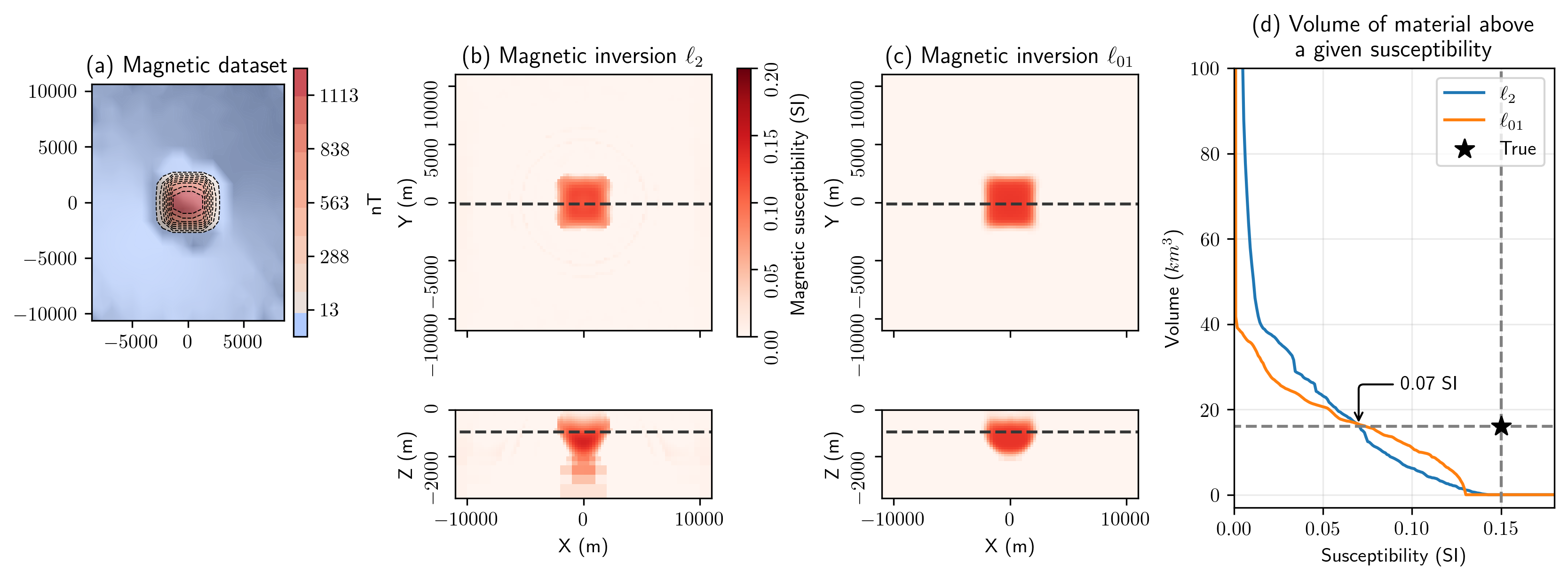}
    \end{center}
\caption{
     (a) Simulated magnetic data for a 4km by 4km block, (b) recovered $\ell_2$ susceptibility model, (c) recovered $\ell_{01}$ susceptibility model, (d) estimated volume of serpentinized rock as a function of threshold choice.
}
\label{fig:one-block-magnetics}
\end{figure}

In Figure \ref{fig:magnetics-l2-lpq-volume}, we see that the threshold that gives us the correct volume for both inversions is 0.08 SI. However, without knowing the true volume, it is difficult to estimate a-priori what the correct threshold should be. In order to estimate this quantity we set up a proxy experiment  in which the geology is a serpentinized block in a half-space. We choose a 4km by 4km block that extends from 500m to 1500m depth, and is positioned in the center of the survey area. We assign the same magnetic susceptibility as the true serpentinized unit, 0.15 SI. We use the same survey geometry as before and perform a forward simulation and add noise. We then perform $\ell_2$ and $\ell_{01}$ inversions of these data using identical model regularizations and parameter choices as in our initial inversions.  The results are shown in Figure \ref{fig:one-block-magnetics}. We then ask the question: what threshold value gives us the correct estimated volume for the block? For both inversions, the correct threshold turns out to be $\sim$0.07 SI; this value is in good agreement with the value of 0.08 SI for the three block model in Figure \ref{fig:magnetics-l2-lpq-volume} Using a threshold of 0.07 SI for the three block model gives us estimates of 40 km$^3$ and 43 km$^3$ for the $\ell_2$ and $\ell_{01}$ inversions, respectively. The threshold values obtained by doing this proxy experiment can be expected to vary with the geometry of the model. We tested this by using a 6km by 6km block and it yielded the same result of 0.07 SI. The procedure has some degree of robustness but choosing a different geometry for the proxy block, changing its dimension, altering the depth, or assigning different values for the assumed physical properties would all have altered the outcome.

\subsection{Gravity}

\begin{figure}[!htb]
    \begin{center}
    \includegraphics[width=0.7\textwidth]{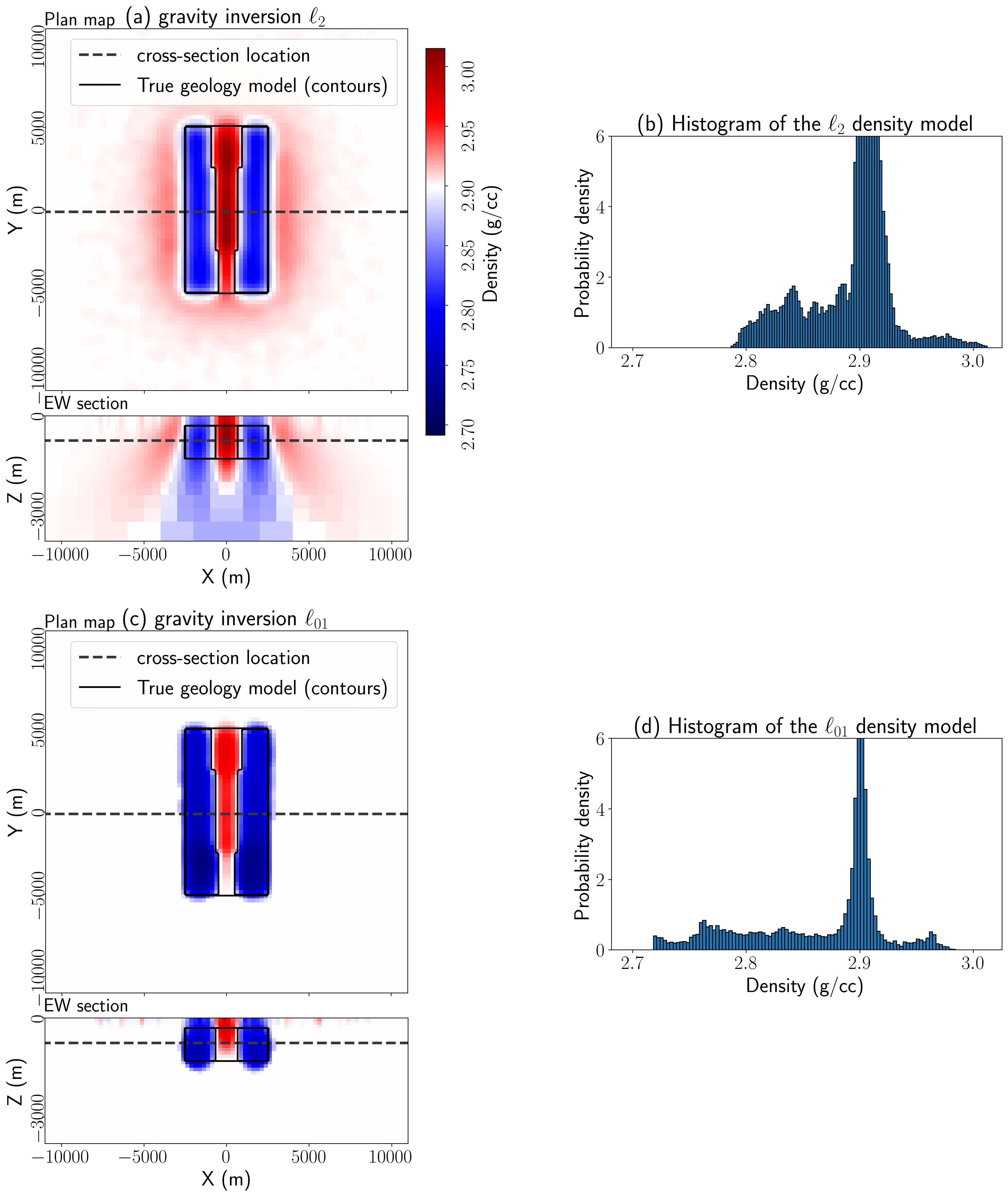}
    \end{center}
\caption{
    Models recovered using (a) an $\ell_2$ inversion of the gravity data and (c) an $\ell_{01}$ inversion of the gravity data. The top plot in each is a depth slice at $z=$-950m and the bottom panel shows a cross section along the line $y=0$m. Plots (b) and (d) show a histogram of the recovered density values.
}
\label{fig:gravity-inversion}
\end{figure}

Next, we perform inversions of the gravity data. Similar to the magnetics, we perform two inversions, first, an $\ell_2$ inversion, shown in Figure \ref{fig:gravity-inversion}(a) as well as an $\ell_{01}$ inversion, shown in Figure \ref{fig:gravity-inversion}(b). Visually, the gravity inversions show a better distinction between the 3 units than the magnetics. Relative to the background, the serpentinized units are less dense, whereas carbonated units are more dense. This is a more diagnostic contrast than what we observed with magnetics. The positive side-lobes we see in the $\ell_2$ inversion result are common when working with blocky models \citep{Oldenburg2005}. The $\ell_{01}$ produces more compact structures, as expected, and reduces the side-lobe effects. It also improves the estimate of the depth of the bottom of the serpentinized unit. Both inversion results show structure coming all of the way to the surface.

\begin{figure}
    \begin{center}
    \includegraphics[width=0.5\textwidth]{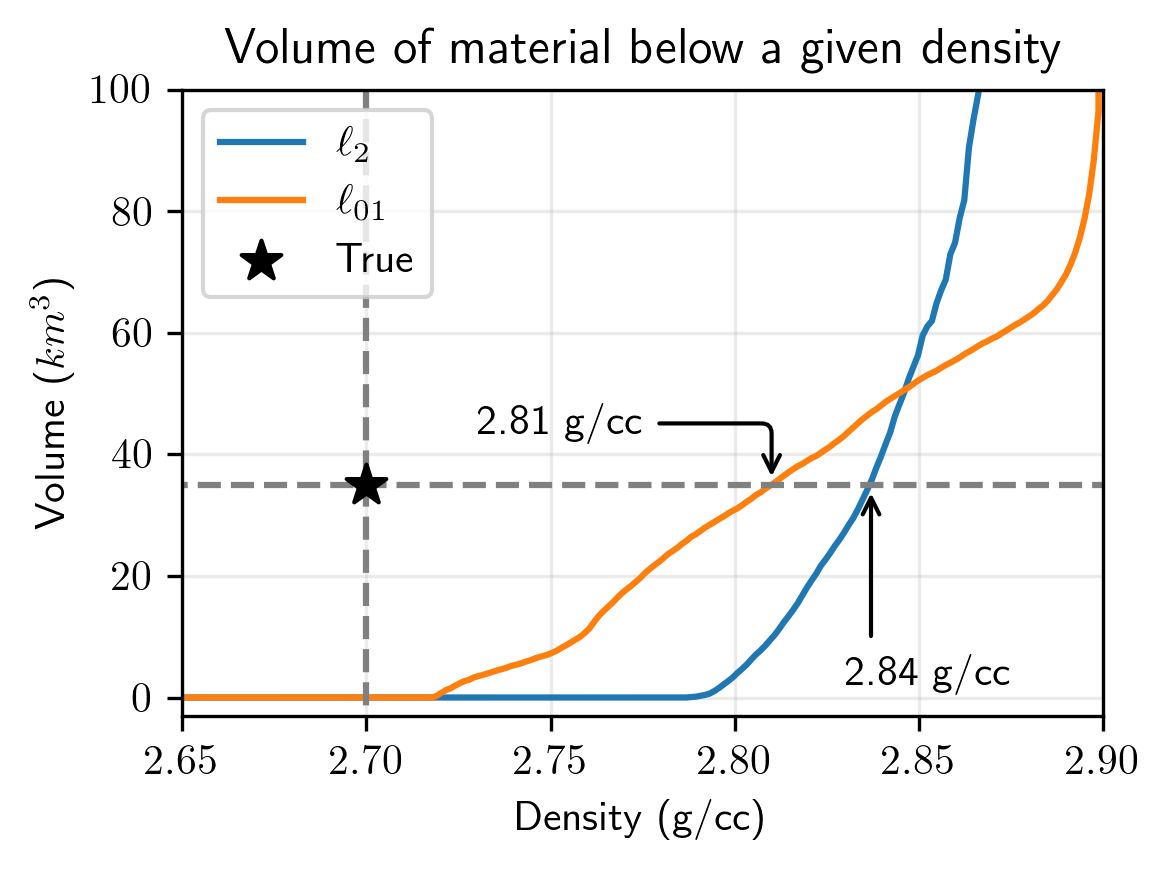}
    \end{center}
\caption{
    Estimated volume of serpentinized rock for a range of density thresholds.
}
\label{fig:gravity-l2-lpq-volume}
\end{figure}

In Figure \ref{fig:gravity-l2-lpq-volume}, we plot the estimated volume of serpentinized rock over a range of density cut-offs. Instead of considering all material above a given threshold as we did in magnetics, we consider all material below a given threshold because the serpentinized rock is less dense than the background. As compared to the magnetics, we can see a greater distinction between the $\ell_2$ and $\ell_{01}$ inversion approaches. The $\ell_2$ has a steeper slope as compared to the $\ell_{01}$ inversion. This indicates that a threshold chosen from the $\ell_{01}$ inversion has a less substantial impact on the estimated volume than the $\ell_2$ inversion.

\begin{figure}[!htb]
    \begin{center}
    \includegraphics[width=1\textwidth]{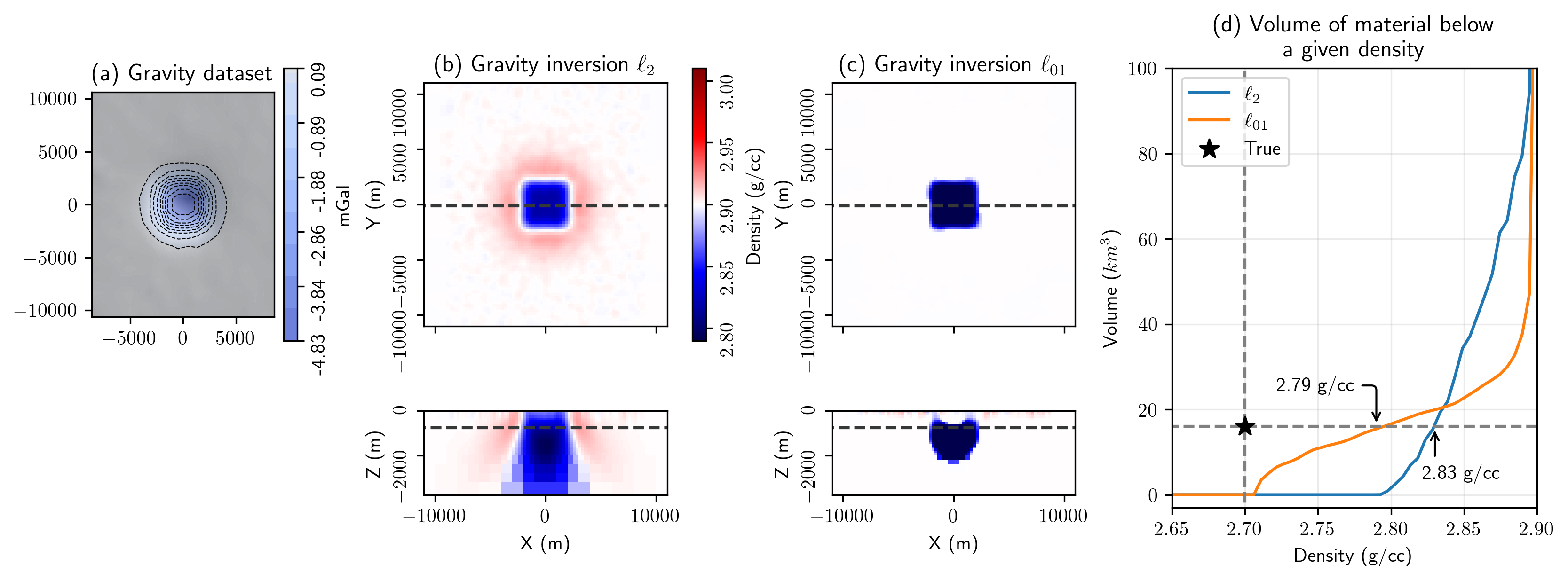}
    \end{center}
\caption{
    (a) Simulated density data for a 4km by 4km block, (b) recovered $\ell_2$ density model, (c) recovered $\ell_{01}$ density model, (d) estimated volume of serpentinized rock as a function of threshold choice.
}
\label{fig:one-block-gravity}
\end{figure}

To estimate the choice of an appropriate threshold, we again perform inversions for the simple 1-block models where the blocks have a density equal to that of the serpentinized unit: 2.7 g/cc. The results are shown in Figure \ref{fig:one-block-gravity}. Similar to the 3 block model, we again see that the volume estimates from the $\ell_2$ inversions have a much steeper slope than the $\ell_{01}$. In this case, the threshold values for the two different regularizations are different:  2.83 g/cc for the $\ell_2$ inversion and 2.79 g/cc for the $\ell_{01}$. This agrees well with the best threshold choices for the three-block model of 2.84 g/cc and 2.81 g/cc for the $\ell_2$ and $\ell_{01}$ results in Figure \ref{fig:gravity-l2-lpq-volume}. Applying thresholds of 2.83 g/cc and 2.79 g/cc to their respective inversions gives us the same volume estimate of 27 km$^3$ for both. The good agreement with choice of threshold and resultant volume estimate supports our hypothesis that our proxy strategy is working so long as we keep all of the parameters in the inversion the same.

\subsection{Summary: Individual Inversions}

Standard $\ell_2$ inversions produce models with smooth structures. For both the gravity and magnetic inversions, we can see this tends to smear out structures in depth. The resultant distribution of physical properties also tends to be smoothly varying with a concentration of values near the background value. Replacing the $\ell_2$ model norm with an $\ell_{01}$ norm promotes more compact structures. In this example, it results in more easily interpretable structures. To estimate the appropriate physical property threshold, we set up a proxy experiment and inverted synthetic data over a simple one-block model. In Table \ref{tab:volume-estimates-summary}, we summarize the estimated thresholds obtained from our proxy-experiment and the resultant estimates of the volume of serpentinized rock from each of the inverted models. The good agreement between both the thresholds and resultant volumes is encouraging for the use of a proxy model for estimating an appropriate physical property threshold to delineate a rock unit. In an analysis of field data, one might run several examples, varying the size and physical properties of a representative target to obtain an estimate and gauge the uncertainty of that estimate.

\begin{table}
    \centering
        \caption{Summary of the thresholds and volume estimates obtained using a proxy experiment of a one-block model. Note that the true volume of serpentinized rock is 35km$^3$}
        \begin{tabular}[htb]{| p{2.5cm} | p{3.25cm} | p{3.25cm} | p{4.5cm} |}
            \hline
            \textbf{Inversion} & \textbf{Threshold for correct volume} & \textbf{Threshold from proxy} & \textbf{Volume estimate with proxy threshold} \\
            \hline
            $\ell_2$  magnetics & 0.08 SI & 0.07 SI & 40 km$^3$ \\
            $\ell_{01}$ magnetics & 0.08 SI & 0.07 SI & 43 km$^3$ \\
            \hline
            $\ell_2$  gravity & 2.84 g/cc & 2.83 g/cc & 27 km$^3$ \\
            $\ell_{01}$ gravity & 2.81 g/cc & 2.79 g/cc & 27 km$^3$ \\
            \hline
        \end{tabular}
        \label{tab:volume-estimates-summary}
     \end{table}

In addition to estimating volumes, the structure and depths of each unit is of practical importance. If serpentinized rock is near the surface, then it may be a candidate for ex-situ carbon mineralization, whereas if it is deeper it is unlikely that it would be economic to mine, but may be a candidate for in-situ carbon mineralization. \cite{Mitchinson2020} uses a depth cutoff of 1km for rocks that may be feasible for ex-situ operations, and within 2km of the surface for in-situ operations. The magnetics inversion shows a good estimate to the depth of the top of the unit, but the gravity brings structure to the surface. So we are left with a question of how to reconcile these results. In the next section, we will examine the use of a Petrophysically and Geologically Guided (PGI) framework for jointly inverting these data.

\section{Including petrophysical data in the inversion}
\subsection{The Petrophysically and Geologically guided Inversions (PGI) framework}

The inverse problem is non-unique. Including knowledge of physical property and geologic information in an inversion can improve our ability to recover structure of interest. Prior information is also a key component to design coupling terms that link together various physical property models, which enables joint inversion of multiple geophysical surveys. The  Petrophysically and Geologically guided Inversion (PGI) framework gives us an avenue for including petrophysical information and jointly inverting multiple geophysical data sets.

\begin{figure}
    \begin{center}
    \includegraphics[width=0.75\textwidth]{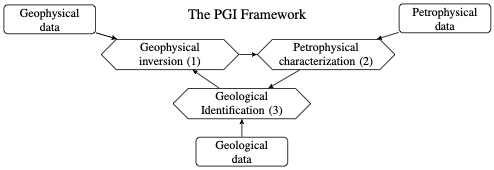}
    \end{center}
\caption{
    The PGI framework (modified from \citet{Astic2019}) is composed of three interlocked inverse problems (diamond nodes) over the geophysical, petrophysical, and geological information (represented by the rectangular nodes).
}
\label{fig:pgi-framework}
\end{figure}

The PGI framework, introduced in \citep{Astic2019} and \citep{Astic2020}, formulates the inverse problem from a probabilistic perspective. Petrophysical and geological information are represented as probability distributions, which are then encoded in the regularization of the Tikhonov objective function (equation \ref{eq:inverse-problem}). To fit all the information at once, the framework cyclically solves three interlocking inverse problems over the geophysical, petrophysical, and geological data (Figure \ref{fig:pgi-framework}). After each geophysical inversion iteration (Figure \ref{fig:pgi-framework}, process (1)), the PGI framework learns a petrophysical representation of the subsurface from the petrophysical data (laboratory measurements, borehole logs, etc.) and current geophysical model (Figure \ref{fig:pgi-framework}, process (2)). The petrophysical model can either be held fixed if we have a strong confidence in the prior petrophysical data or it can be iteratively updated to benefit from the knowledge gained at each iteration from the geophysical data. Then, PGI builds a quasi-geology model \citep{Li2019}, a representation of the geology based on the geophysics, by classifying the current geophysical model into rock units using the petrophysical model and geological information. This quasi-geology model serves to update the regularization weights and reference model for the next geophysical inversion iteration; the cycle continues until all geophysical, petrophysical, and geological data are fit. More details about these different processes are given in \citep{Astic2019}.

\subsubsection{Representation of petrophysical information: Gaussian Mixture Model (GMM)}

\begin{figure}
    \begin{center}
    \includegraphics[width=0.5\textwidth]{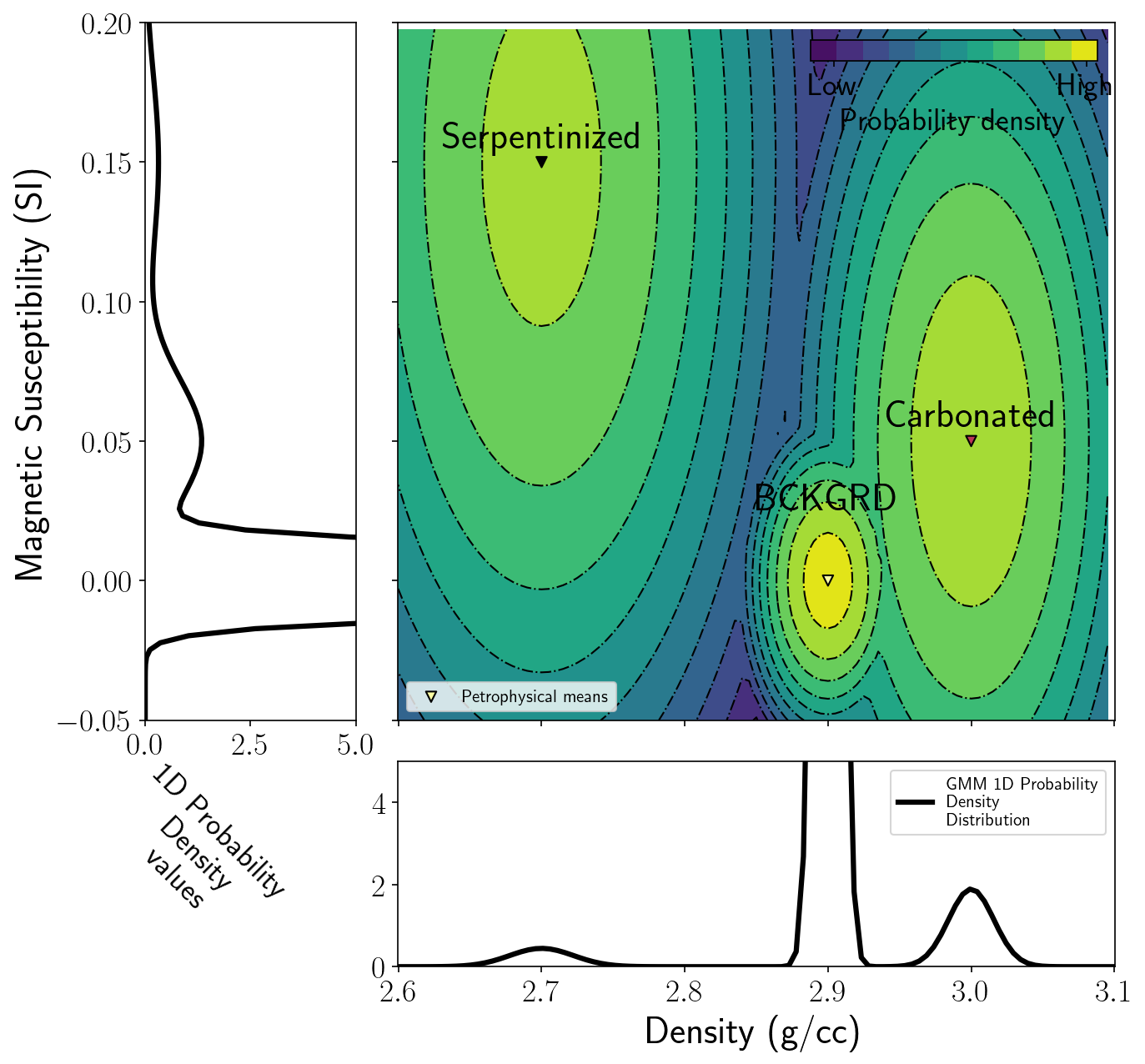}
    \end{center}
\caption{
    Physical properties information represented as a GMM. The left and bottom panels show the magnetic susceptibility and density probability distribution, respectively. The middle panel represents their joint probability distribution.
}
\label{fig:synthetic-gmm}
\end{figure}

Within the PGI framework, petrophysical information is represented as a probability distribution modelled by a Gaussian Mixture model (GMM)  (example in Figure \ref{fig:synthetic-gmm}). A GMM is composed of a weighted sum of normal distributions, also called Gaussians. Each Gaussian distribution represents a rock unit. The goal of a PGI is to find a physical properties model of the underground that both fits the geophysical data and reproduces the GMM distribution.

In the present case, we have three rock units: the background, the carbonated, and the serpentinized facies. Each formation is characterized by: (1) the mean of its physical properties (for example 3.0 g/cc and 0.05 SI for the carbonated; represented as triangles in Figure \ref{fig:synthetic-gmm}), (2) its covariance which quantifies how spread its petrophysical values are around the mean, and (3) the probability of each rock prior to geophysical observation. This last parameter weights each normal distribution and can encode local geological information, or just be set at a constant global value everywhere if no such information is available. The chosen global value has minimum impact on the inversion \citep{AsticThesis}. In Figure \ref{fig:synthetic-gmm}, the central panel presents the joint density and magnetic susceptibility contrasts probability distribution; this is what is used to perform multi-physics inversion with PGI, using the petrophysical distribution as a coupling term. The side panels represent the ``marginal probability distribution'' for each physical property individually; they will be used for the individual magnetic and gravity inversions.

\subsection{Individual PGI: Magnetics}
We start by performing a PGI inversion of each data set independently. In addition to requiring that the model fit the geophysical data through the use of a geophysical data misfit, we now also include a petrophysical misfit term. With the geophysical data misfit, we compare the data predicted from a forward simulation over the geophysical model with the true, observed data. With the petrophysical misfit term, we compare the physical property estimated from the geophysical model with the mean physical property of the rock unit it is classified as in the geologic identification. For example, a cell with a value of 0.13 SI would be classified as serpentinized and therefore compared with the mean value of the serpentinized facies (0.15 SI), whereas a cell with a value of 0.06 SI would be classified as carbonated and compared with that value (0.05 SI). To construct the misfit metric, standard deviations of the petrophysical data are also taken into account; these details are discussed in \cite{Astic2019}. PGI is an iterative procedure, so with each update to the physical property model in the geophysical inversion component, we re-evaluate the petrophysical misfit and update the associated quasi-geology model. In doing so, we obtain a model that respects both the geophysical and petrophysical data. Further, the quasi-geology model is constructed as a part of the inversion, meaning we do not need to select a threshold to identify serpentinized rock and estimate its volume, as we did in the previous inversions. Here, we can use the geologic classification produced by PGI and directly estimate the volume from the quasi-geology model.

\begin{figure}[!htb]
    \begin{center}
    \includegraphics[width=0.9\textwidth]{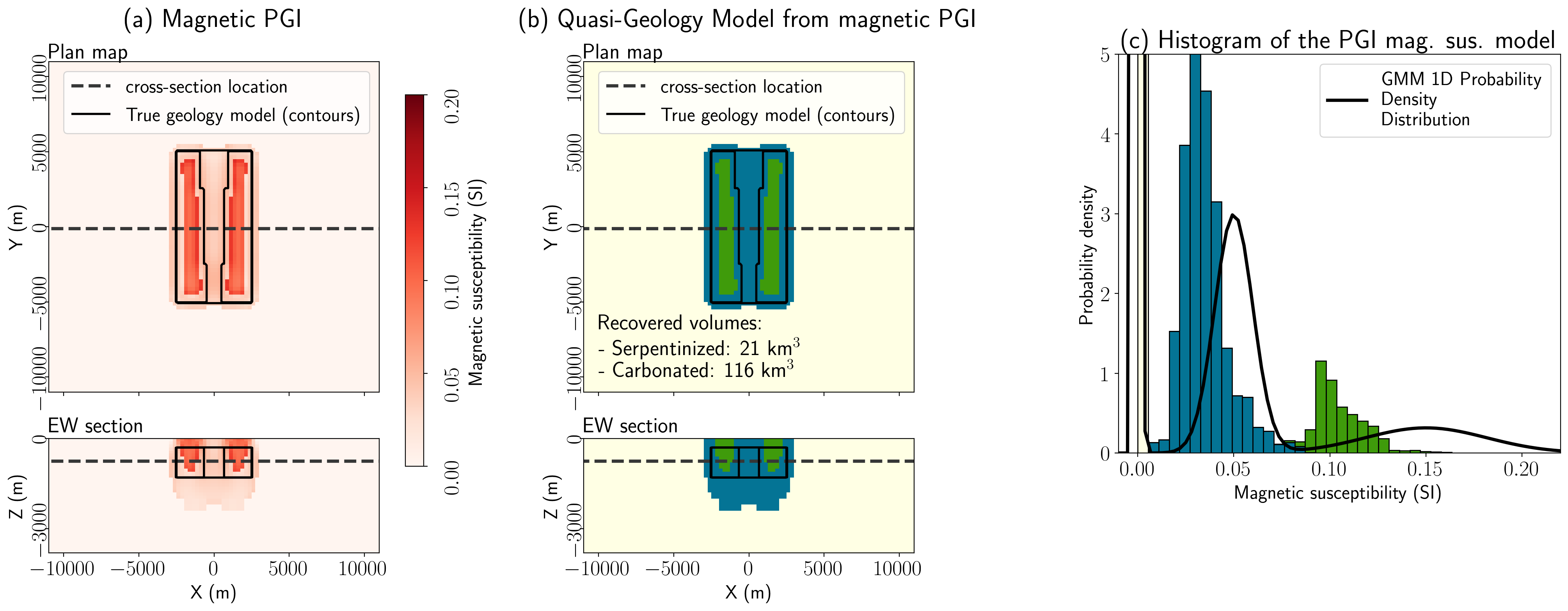}
    \end{center}
\caption{
    Results from inverting the magnetics data with the PGI framework. (a) Magnetic susceptibility model, (b) associated quasi-geology model, and (c) histogram of the recovered magnetic susceptibility values.
}
\label{fig:single-physics-pgi-magnetics}
\end{figure}

In Figure \ref{fig:single-physics-pgi-magnetics}, we show the outcome of PGI applied to the magnetic data. In this case, we see that, we successfully identify two blocks of higher susceptibility whose position and petrophysical signature correspond well to the serpentinized unit in Figure \ref{fig:single-physics-pgi-magnetics}a). Contrary to the $\ell_2$ and $\ell_{01}$ inversions, the histogram in (c) shows distinct clusters that correspond to the petrophysical characteristics of each rock unit (background in white, serpentinized in green, and carbonated in blue). The solid line in (c) shows the GMM we chose for this inversion. The background has a mean susceptibility of 0 SI, the carbonated unit 0.05 SI, and the serpentinized unit 0.15 SI. This is what we compare the physical property values from the geophysical inversion to in the petrophysical misfit. A ``perfect'' inversion would have a histogram that aligns with the defined GMM. In this inversion, the distributions of the recovered susceptibilities for the carbonated and serpentinized unit are shifted to the left, having lower mean values. There are also more values attributed to the carbonated unit than the true distribution. This effect, as well as the ``halo'' structure we observe in the quasi-geology model (b), where the regions classified as serpentinized rock are surrounded by carbonated rock magnetic susceptibility, can occur in single physical property inversions. When two units have contrasts that are both positive or both negative compared to the starting value of the background unit, the inversion tries to explain the data first with smaller contrast (carbonated rock) and only puts larger values where it is strongly indicated in the data (see \citet{AsticThesis} for more information on ``consecutive clusters''). As a result, we underestimate the volume of serpentinized rock, at 21 km$^3$. With PGI, we are also able to estimate the volume of carbonated rock (116 km$^3$). This would be challenging to do with the $\ell_2$ and $\ell_{01}$ inversions as it would require that we choose both upper and lower bounds to attribute to the carbonated rock.

\subsection{Individual PGI: Gravity}
\begin{figure}[!htb]
    \begin{center}
    \includegraphics[width=0.9\textwidth]{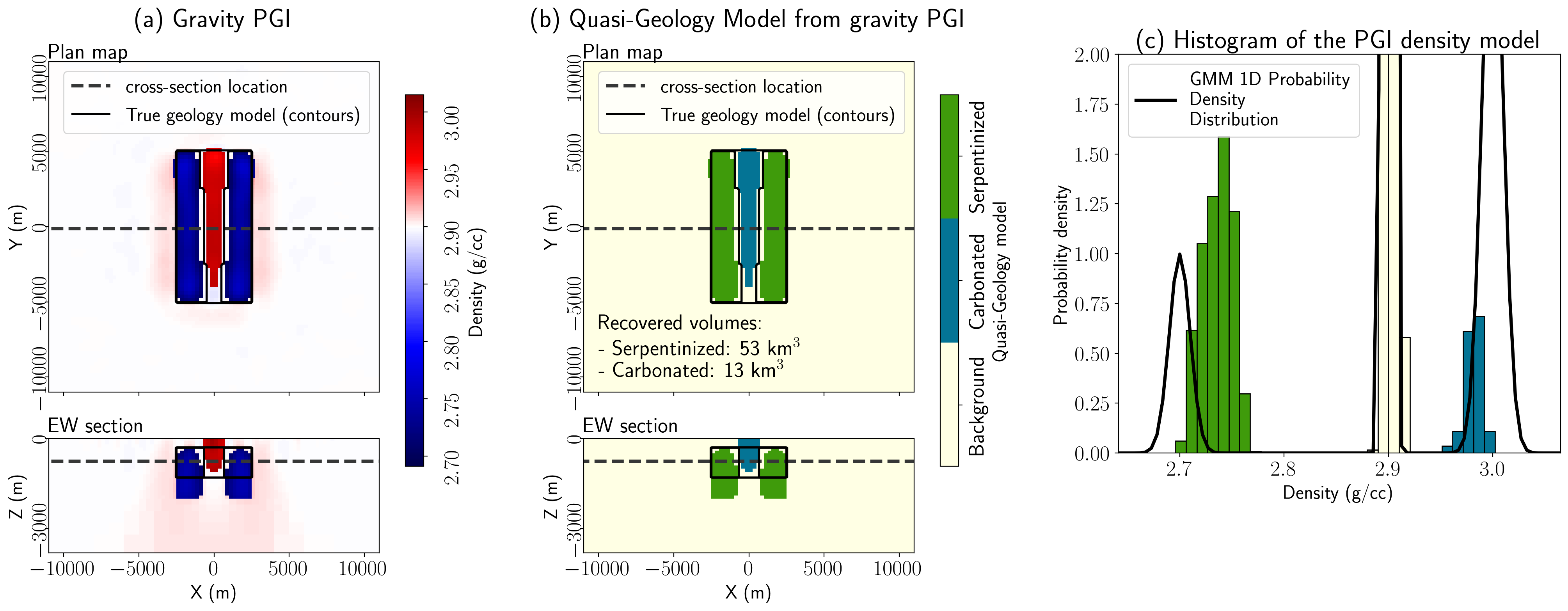}
    \end{center}
\caption{
    Results from inverting the gravity data with the PGI framework. (a) Density model, (b) associated quasi-geology model, and (c) histogram of the recovered density values.
}
\label{fig:single-physics-pgi-gravity}
\end{figure}

Next, we perform PGI using the gravity data and show the results in Figure \ref{fig:single-physics-pgi-gravity}. We see that 3 distinct volumes are recovered, two with the petrophysical signature of the serpentinized unit (mean density of $\sim2.7$ g/cc), and one in the middle with the petrophysical signature of the carbonated unit (mean density of $\sim30$ g/cc). Since the physical property contrasts are opposite in sign (the density of the serpentinized unit is less than the background, whereas the density of the carbonated unit is larger), this is a simpler distribution to work with in the PGI framework than we encountered with magnetics. In the histogram in Figure \ref{fig:single-physics-pgi-gravity} (c), we see that the mean contrasts are slightly underestimated, but adequately fit, and there is a  clear separation between the physical properties of each unit. There are also improvements in the recovered model. The depth to the top of the unit identified to be serpentinized is well-recovered, in contrast to what we observed in the $\ell_2$ and $\ell_{01}$ inversions (Figure \ref{fig:gravity-inversion}). The total volume of serpentinized rock is estimated from the quasi-geologic model, 53 km$^3$ for the serpentinized unit and 13 km$^3$ for the carbonated unit. The total volume of carbonated rock is quite close to the true volume (15 km$^3$), and the volume of serpentinized rock is overestimated, which may be expected because the recovered densities of the serpentinized rock has a lower contrast than the true value.

There are several benefits to the PGI framework for this application: (a) prior petrophysical knowledge can be included in the inversion (with uncertainties); (b) rather than a smooth range of physical property values recovered in the $\ell_2$, $\ell_{01}$ inversions, the recovered petrophysical signatures are more representative of the true petrophysical signature; and (c) the quasi-geology model, built by reproducing the petrophysical signature, gives a volume estimate that is independent of the choice of an ad hoc threshold.

Despite the listed benefits in using a PGI approach, we note that the quasi-geology models obtained from magnetics and gravity are different. For example, in the magnetics inversion, the serpentinized unit comes to the surface whereas in the gravity inversion, the depth to the top of this unit is $\sim$ 300m. Also the  volume estimates of the serpentinized body are different and neither is in agreement with the true model. The differences between the two quasi-geology models can be reconciled by taking advantage of the complementary information in gravity and magnetic data. The next step is to use PGI to jointly invert these data sets.

\subsection{Joint PGI}
Now, we seek to find a model that fits both geophysical data sets and is consistent with our knowledge of the petrophysics. There are many avenues for approaching a joint inversion, and \cite{Haber2013, Moorkamp2016} provide an overview of some approaches. The two major categories use either structural or petrophysical characteristics to couple data that are sensitive to different physical properties. The PGI approach uses petrophysical information to couple multiple data sets. Now rather than the marginal distributions, we will work with the 2D GMM as shown in Figure \ref{fig:synthetic-gmm}. For the joint PGI, we are asking the inversion now to fit: a misfit associated with the magnetics data, a misfit associated with the gravity data, and a petrophysical misfit that is defined in 2D space. The petrophysical misfit is what couples the two data sets, and in looking at Figure \ref{fig:synthetic-gmm}, what is considered for that misfit is now distances in 2D space from the cluster centers. All of these misfit terms must be balanced appropriately, and we provide details on our approach in \citep{Astic2020}.

\begin{figure}[!htb]
    \begin{center}
    \includegraphics[width=0.75\textwidth]{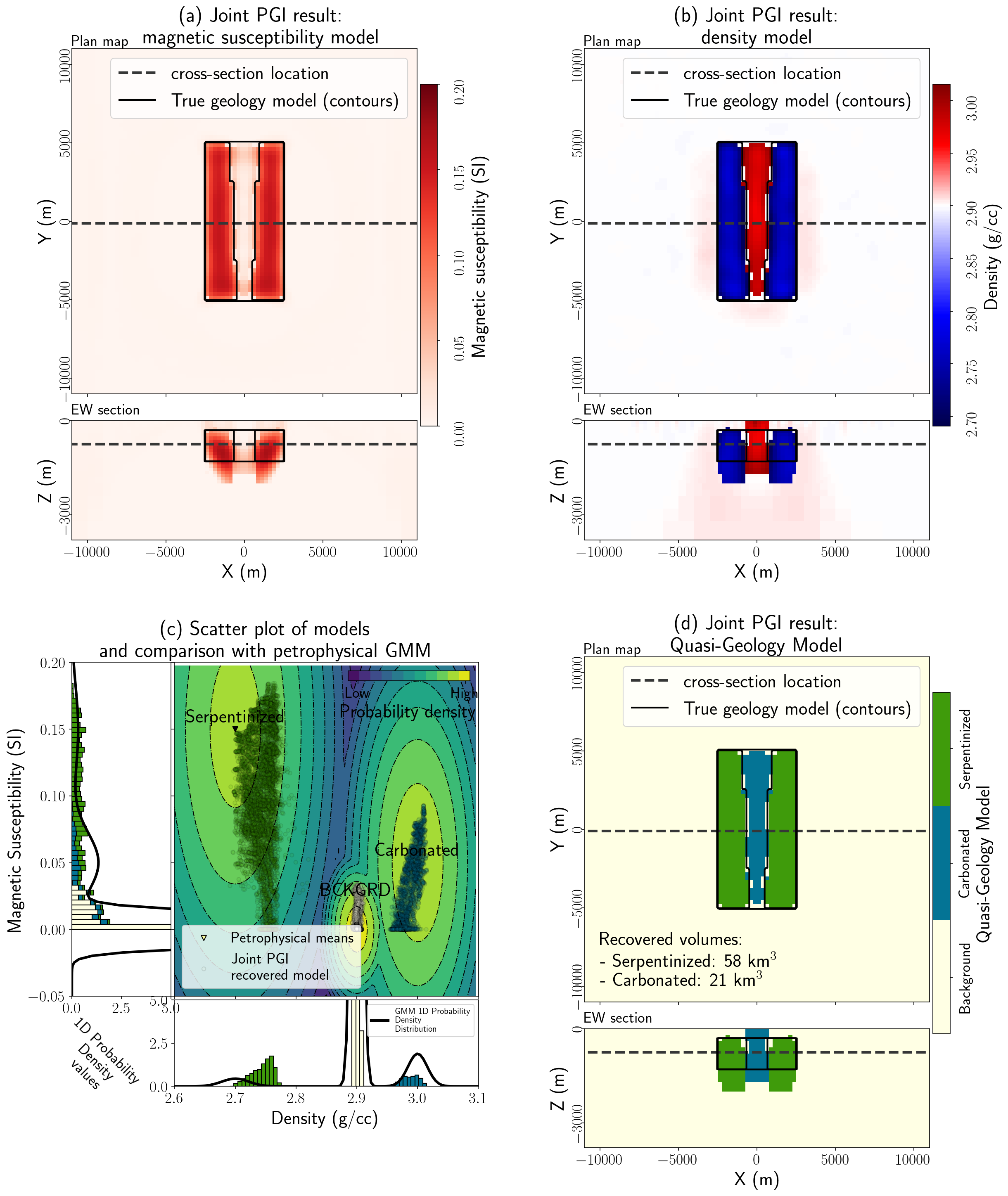}
    \end{center}
\caption{
    Results from jointly inverting the magnetics and gravity data with the PGI framework.
}
\label{fig:joint-pgi}
\end{figure}

Using the PGI framework, we perform a joint inversion with the gravity and magnetic surveys. The result of that joint inversion can be seen in Fig \ref{fig:joint-pgi}. First, examining the magnetic susceptibility model, we see that the depth to the highly susceptible serpentinized unit is in much better agreement with the true model, and we are no longer plagued by the ``halo'' effects we saw in Figure \ref{fig:single-physics-pgi-magnetics}. Looking at the density model first (Fig \ref{fig:joint-pgi}a), we see that the contacts between the various units are much better recovered thanks to the addition of the magnetic information. Both models share the same underlying quasi-geology model (Fig \ref{fig:joint-pgi}c). The petrophysical information is fit (Fig \ref{fig:joint-pgi}d, where we show the coupling GMM and the crossplots of the physical properties models), it appears that the density signatures are better reproduced than in the individual PGI inversion. There is more spread in the distribution of magnetic susceptibility than the density values, but this is reflective of what is expected (e.g. see the GMM in Figure \ref{fig:joint-pgi}c), and by using multiple physical properties in the inversion, we can better handle this ambiguity. Using the joint PGI approach, we produce a single quasi-geology model that is consistent with the magnetics and gravity data, as well as our prior knowledge of the petrophysics. The total volume estimates of the serpentinized and carbonated rock units are 58 km$^3$ and 21 km$^3$, respectively.

\section{Cumulative volume estimates}
It is satisfying to obtain a result that is consistent between the magnetics and gravity data and overall does a good job delineating the three rock units as we did with the joint PGI inversion (Figure \ref{fig:joint-pgi}). The total estimated volume of serpentinized rock is overestimated, which is due to the extra material at depth, beneath the unit. Both gravity and magnetic methods have limited depth information. Sensitivity decreases with depth and at some point our recovered models reflect a-priori information and the nature of our algorithm. Thus, our estimates of total volume can be impacted by this material at depth which may not be substantially supported in the data. As mentioned previously, the total volume is of interest, but perhaps more important is the volume as a function of depth as this controls whether it is practical to perform ex-situ or in-situ carbon mineralization operations.

\begin{figure}
    \begin{center}
    \includegraphics[width=0.9\textwidth]{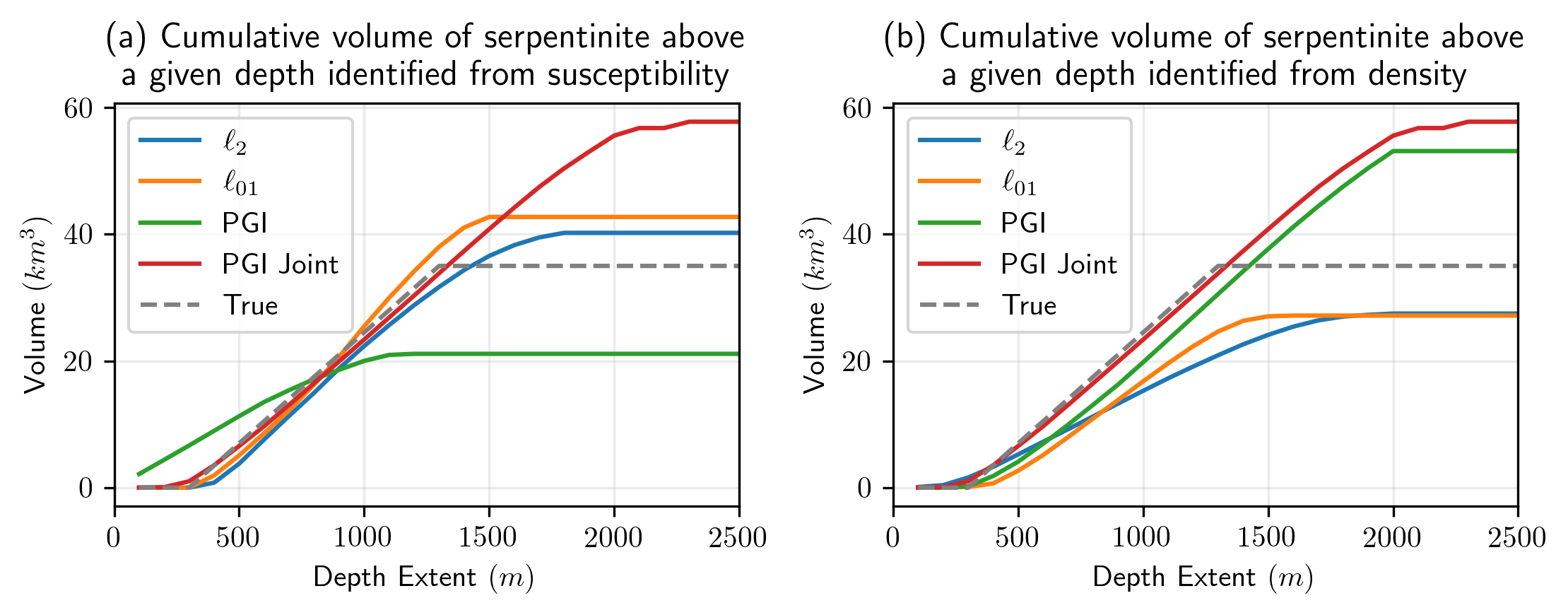}
    \end{center}
\caption{
    Estimated volume of serpentinized rock above a given depth for the (a) susceptibility and (b) density models recovered in each of the inversions.
}
\label{fig:volume-depth}
\end{figure}

To compare the various approaches, we plot the cumulative volume of serpentinized rock estimated from each of the inversions as a function of depth in Figure \ref{fig:volume-depth}. Depth extent is on the x-axis, and the volume of rock above a given depth that we attribute to serpentinized rock from each inversion is on the y-axis. For the $\ell_2$ and $\ell_{01}$ inversions, we use the threshold values from the 4km block as defined in table \ref{tab:volume-estimates-summary}. For the volume estimates as a function of depth from magnetics, we see that overall, there is good agreement between the inversion results, with the exception of the individual PGI. The poor estimation with depth from the PGI inversion of the magnetics data alone is consistent with the cross section observed in Figure \ref{fig:single-physics-pgi-magnetics}. More of the susceptible material was attributed to the carbonated unit than the serpentinized unit. This highlights the value of running multiple styles of inversions. Where there is consistency between results, we can be more confident in the resultant model. Turning our attention to the gravity, we see that the $\ell_2$ inversion puts material right at the surface and both the $\ell_2$ and the $\ell_{01}$ underestimate the volume as a function of depth at 1000m depth. The individual gravity PGI is an improvement. It slightly underestimates the volume between 500 and 1300m depth (the true depth of the bottom of the unit). The joint PGI models result in the best agreement between the surface and 1300 m depth. It also has the advantage that it is consistent for both susceptibility and density, because we require that the inversion produces a result that agrees with both data sets.

\section{Summary}
Carbon mineralization could be an important mechanism for reducing atmospheric CO$_2$ as we work to transition to a carbon neutral economy. When rocks with CO$_2$ sequestration potential have physical properties that are distinct, geophysical data can be used to locate and delineate rocks with sequestration potential. Inversion is an essential tool for obtaining models of the subsurface that we can interpret geologically. The inverse problem is non-unique, and therefore assumptions and prior information need to be included. Defining a model norm is a standard approach. Typically an $\ell_2$ norm is used, and this tends to produce models with smooth features. More compact structures can be promoted by using norms less than 2. Using an $\ell_{01}$ produced more compact structures. This can simplify the interpretation of the resultant model because boundaries between units are more easily defined, and some inversion artefacts were reduced. When estimating a volume of rock, which is an important factor for the amount of CO$_2$ that can be sequestered, the use of an $\ell_{01}$ norm produced estimates that were less sensitive to the exact choice of threshold than an $\ell_2$ inversion.

Using the Petrophysically and Geologically guided Inversion (PGI) framework, we illustrated how knowledge of physical properties can be brought into the inversion. There are several benefits that this inversion approach offers: (1) physical property information, with associated uncertainties, is fit as a part of the inversion, (2) a quasi-geology model is produced, and (3) joint inversions that use multiple data sets that are sensitive to different physical properties can be performed. By using the joint PGI framework, we produced a quasi-geology model that was consistent with both the gravity and magnetics data sets. The model that was produced was also the best for estimating the volume of rocks with sequestration potential as a function of depth.

The inversions employed, the $\ell_2$, $\ell_{01}$, and the PGI, increase in complexity and the user-knowledge that is required to obtain reasonable results. The more complex inversions can offer some benefits, as we have described. This is not to say that they should be the first inversion employed. In general, starting with the simpler $\ell_2$ inversion is advisable for a first interpretation. It is much simpler to run $\ell_2$ inversions to examine the impacts of choices of inversion parameters such as the noise characteristics and the use of depth or sensitivity weighting. Once these have been tuned on a simple inversion, they can be used for the more complex inversions. It is also advantageous to have multiple models of the subsurface to examine which features are consistent, or not between them.

The model we chose to work with was seemingly quite simple, but shows that inversion, and obtaining high quality models of the subsurface is an iterative process. We did not include the impacts of remanent magnetization in our analysis. This will be important to address and presents an opportunity for future research.

\section{Acknowledgements}
The authors are grateful to Greg Dipple, Peter Bradshaw, and Randy Enkin for multiple conversations about the geochemistry and important questions for carbon mineralization. We are grateful to Dianne Mitchinson for discussions as we designed a representative model and Jamie Cutts for making the physical property data available. Finally, we are grateful to the SimPEG team, and Dominique Fournier in particular for making his work on compact and sparse norms available through the project.

\clearpage
\bibliographystyle{seg}
\bibliography{fasttimes-carbon-mineralization}

\end{document}